\documentclass[amsfonts, 10pt]{article}
\usepackage{amssymb}
\newtheorem{theorem}{Theorem}
\newtheorem{definition}{Definition}

\newtheorem{lemma}{Lemma}
\title{The Born rule from a consistency requirement on hidden measurements in complex Hilbert space}
\author{Sven Aerts\footnote{CLEA, Vrije Universiteit Brussel (VUB), Krijgskundestraat 33, B-1050 Brussels, Belgium, electronic address: saerts@vub.ac.be}}
\date{}
\begin{document}
\maketitle

\begin{abstract}
We formalize the hidden measurement approach within the very general notion of an interactive probability model. We narrow down the model by assuming 
the state space of a physical entity is a complex Hilbert space and introduce the principle of consistent interaction which effectively partitions the space of apparatus states. 
The normalized measure of the set of apparatus states that interact with a pure state giving rise to a fixed outcome is shown to be in accordance with the probability 
obtained using the Born rule.
\end{abstract}

\section{Introduction}

In \cite{Aerts1986}, Aerts D. outlines a proposal to answer the question of the arisal of
probabilities in quantum mechanics. The author argues that
probability enters quantum mechanics because of a lack of knowledge about
which measurement was conducted. Let us briefly outline the scheme as presented in the article 
to reproduce the probabilities related to the measurement 
of an observable $\mathcal{A}$  with $n$ possible alternative (and mutually exclusive) outcomes. 
The $n$ eigenvectors $\{e_1,e_2,\ldots,e_n\}$ of the operator $A$ that represents the
observable $\mathcal{A}$  with $n$ possible outcomes, can serve as a basis
for the state of the entity: $ q=\sum_{i=1}^n\langle q,e_i\rangle e_i $. Orthodox quantum mechanics dictates that the probability 
$p_q^{\mathcal{A}}(a_i)$ of finding the result $a_i$ -one of the eigenvalues 
$\{a_1,a_2,\ldots,a_n\}$ of the eigenvector with same index- upon execution of
the measurement that corresponds to the observable $\mathcal{A}$ when the
entity is in the state $q$, equals 
$$p_q^{\mathcal{A}}(a_i) \equiv p({\mathcal{A}}= a_i | q)=| \langle q,e_i\rangle| ^2 $$
This means that the n-tuple 
$$ \kappa =(p_q^{\mathcal{A}}(a_1),p_q^{\mathcal{A}}(a_2),\ldots,p_q^{\mathcal{A}}(a_n))$$
contains all statistical information we can derive from the entity with
respect to the observable $\mathcal{A}$ and as such the author argues, we
can use $\kappa $ as a representation of the statistical state. Because the 
$ p_q^{\mathcal{A}}(a_i)$ are constrained by the requirement 
$\sum_{i=1}^np_q^{\mathcal{A}}(a_i)=1,$ we see that the statistical state is an element of the 
$(n-1)$-simplex $\Delta _{n-1}$ in $\mathbb{R}^n$ spanned by the canonical base
vectors $e_i:$ $\kappa =\sum_ip_q^{\mathcal{A}}(a_i)e_i$ . The basic idea of
the ''hidden measurement approach'' is to associate with each measurement $m$
a set of sub-measurements $m(\lambda )$ such that the measurement $m(\lambda)$ 
consists of choosing at random one of the $\lambda $ and performing the
measurement $m(\lambda )$ on the entity. The measurements are to be taken
classically deterministic, in the sense that their operation on a fixed
state always yields the same result. This is done as follows: take $\lambda $
to be an $n$ -tuple from the $(n-1)$-simplex: $\lambda =(\lambda _1,\ldots,\lambda _n),\sum \lambda _i=1,\lambda _i\geq 0 $.
Call $C_i$ the convex closure of the set 
$\{e_1,..,e_{i-1},\kappa ,e_{i+1},\ldots,e_n\}.$ The outcome of the measurement $m(\lambda )$ is
determined by $\lambda $ in the following \emph{ad hoc} way: if 
$\lambda \in C_i$ , then the outcome reads $a_i.$ We will not discuss the procedure when
the variable $\lambda $ happens to be chosen on the boundary of one of the
$C_i$ as this is a thin subset only and as such does not
contribute to the final probabilities. The probability of choosing $\lambda$
in the simplex $C_i$ is calculated by assuming a uniform and normalized
density for $\lambda $ over $\Delta _{n-1}.$ Hence we obtain 
$$p(\lambda \in C_i|q)=\mu (C_i)/\mu (\Delta _{n-1}) $$
where $\mu $, because of the uniform measure, is simply the $(n-1)$-dimensional volume of the
respective simplex. This volume is proportional to both the measure of any
of its $(n-2)$-dimensional faces and to the length of the orthogonal
projection of its ''height'' onto this face. Hence we can easily see that
the volumes of the simplices are proportional to the projections of the
statistical state $\kappa $ onto the base vectors. It is a matter of
straightforward determinant calculus to show that 
$\mu (C_i)/\mu (\Delta_{n-1})=p_q^{\mathcal{A}} (a_i),$ and hence we have: 
$$p({\mathcal{A}}=a_i | q)=p(\lambda \in C_i | q) $$ 
The result is deceivingly simple and it is difficult to imagine a shorter
exposition of the well-known fact that there exist hidden variable models of
quantum mechanics if one restricts the latter to measurements related to a
single observable. This strength is immediately also a weakness of the
exposition: the state of the entity is identified with the statistical
state, or the set of probabilities related to a single observable, whereas
in quantum mechanics we are able -at least in principle- to apply Dirac
transformations to calculate the probabilities related to \emph{all}
observables we choose to measure. It is not obvious how to transform
the state in the simplex when we want to measure a different observable. Is
it possible to extend the procedure and make it work in Hilbert space rather
than in the simplex? In the original article such an example is indeed
given, but it relates only to a two-dimensional problem. However, the two
dimensional case is in some sense a degenerate case: the possibility 
of sub-measurements is excluded and the Gleason theorem applies only from
dimension three or higher. The latter fact has sometimes been related to the
existence of hidden variable models for measurements with only two outcomes.
To counter this objection, a three-dimensional model in a real Hilbert space
\cite{3dsphere} was constructed. However, this model is
much more complicated and ad-hoc than the original model and did not give a
hint as to how and if the scheme would work in complex Hilbert space.
Although a model in complex Hilbert space was lacking, interesting results 
in other directions where obtained. For example, the
question of the generality of the measure theoretic construct was adequately
dealt with in a lattice-theoretic model for an experiment with possibly
infinite outcomes \cite{coecke}. The two dimensional model easily allowed for
parametrization of the lack of knowledge, engaging us to study the behavior of
between quantum and classical descriptions by means of
statistical polytopes and the violation of the axioms of quantum logic. 
We refer to \cite{Aerts1999} \cite{inter1} \cite{inter2} and the references found there.

\noindent The present article aims at resolving two issues. The
first one, raised at the end of the 1986 paper, is how to characterize the
measurements that occur in a hidden measurement scheme. Can we give a less
ad-hoc description of the way a measurement selects an outcome when it 
interacts with a state? The second issue is concerned with the realization of
 such a scheme in complex Hilbert space. 
More precisely, we will put forward a principle that
partitions the set of measurements such that the measure of the set of
apparatus states that actualize a fixed outcome if in interaction with a
system in a pure state, is shown to be equal to the modulus squared of the
inner product of the state of the entity with the eigenstate belonging to
that particular outcome. 

\section{Lack of Knowledge in an Interactive set-up}

We will first recast the hidden measurement idea into the more general and 
abstract notion of an interactive system. In essence, we assume the observer
is in a state $a\in M$, and the thing he observes is in a state 
$q\in \Sigma.$ Furthermore we assume the existence of a rule of interaction $``i"$ that
gives us the outcome $x\in X$ as a result of the interaction between the two
states $q$ and $a$: 
$$i:\Sigma \times M\rightarrow X,\ i(q,a)=x$$
We want this model to be deterministic, hence the mapping $i$ is a function.
Furthermore, we want every possible outcome $x$ to be the result of an
interaction between an entity and a measurement apparatus, hence we also
require $i$ to be surjective. Of course, surjectivity implies the
possibility that different couples $(q,a)$ lead to the same outcome: $i^{-1}(x)=\{(q,a)\in \Sigma \times M:i(q,a)=x\}$

\noindent Suppose now that we have a lack of knowledge about the precise state of the
system and apparatus. With ${\mathcal{B}}(\Sigma )$ (and ${\mathcal{B}} (M)$)
the Borel field of $\sigma $-additive subsets of $\Sigma $ (and $M$), our
experiment is characterized by two probability measures: $\mu _\Sigma $ as a
probability measure from ${\mathcal{B}} (\Sigma )\rightarrow [0,1]$ and $\mu _M$
as probability measure ${\mathcal{B}}(M)\rightarrow [0,1]$:
\begin{eqnarray*}
\ {\mathcal{P}}_\Sigma &=&(\Sigma, {\mathcal{B}}(\Sigma ),\mu _\Sigma )
\\ {\mathcal{P}}_M &=&(M,{\mathcal{B}}(M),\mu _M)
\end{eqnarray*}
The way the system and the apparatus interact is goverened solely by the
function $i$: the measures themselves are independent. To define the
probability of the occurrence of an outcome, we assume $i$ is a measurable
function and as such, the interaction $i$ becomes an independent random
variable from $\Sigma \times M$ onto $X$. First we need a few definitions (all sets are assumed to be non-empty):

\begin{definition} 
An \emph{interactive probability model} is a
quadruple $({\mathcal P}_\Sigma, {\mathcal P}_M,X,i)$ with:

\noindent ${\mathcal P}_\Sigma =(\Sigma, {\mathcal B}(\Sigma ),\mu _\Sigma )$, 
a probability space of a set of entity-states $\Sigma$,

\noindent ${\mathcal P}_M=(M, {\mathcal B}(M),\mu _M)$, a probability
space of a set of apparatus-states $M$, 

\noindent a non-empty set $X$  called the outcome space, and 

\noindent a random variable $i:\Sigma \times M\rightarrow X$, called the interaction.
\end{definition}

\begin{definition}
A \emph{preparation} $\pi =(\psi _q,\psi _a)$
is an ensemble of entity states $\psi _q\in {\mathcal B}(\Sigma )\ $
and an ensemble of apparatus states $\psi _a \in {\mathcal B} (M)$.
\end{definition}

\noindent The odds of picking a certain system state out of the ensemble $\psi _q$ and
picking one apparatus state out of $\psi _a,$ is determined independently by
the measures $\mu _\Sigma $ , resp. $\mu _M.$

\noindent Following standard probability theory, we construct the product space 
${\mathcal P}_{\Sigma \times M}=(\Sigma \times M,{\mathcal B}(\Sigma )\times 
{\mathcal B}(M),\rho )$. The measures $\mu _\Sigma $ and $\mu
_M$ induce the unique product probability measure $\rho :$ ${\mathcal B}(\Sigma )\times {\mathcal B}
(M)\rightarrow [0,1]$, such that 
$\rho (\psi _q,\psi _a)=\mu _\Sigma (\psi_q)\mu _M(\psi _a)$. 
This leads to the following definition:

\begin{definition}
Given an interactive probability model 
$({\mathcal P}_\Sigma ,{\mathcal P}_M,X,i)$ and a preparation 
$\pi=(\psi _q,\psi _a)\in {\mathcal B}(\Sigma )\times {\mathcal B}(M)$
The \emph{interactive probability} of the occurrence of the outcome $x$: 

$$p(x\ | \pi \ )=\frac 1{\rho (\psi _q,\psi _a)} \int_{i^{-1}(x)} d\rho $$
\end{definition}

\smallskip 
\noindent We stress that this definition of the probability allows for a
completely natural lack of knowledge interpretation: any arising
probability in the occurrence of outcomes, is a consequence of the inability
to prepare identical states for either the system, the apparatus, or both.
This point is crucial. If we have an irreducible uncertainty about the way
we study nature, it will be impossible to give a direct operational meaning
to both $(\Sigma ,{\mathcal B}(\Sigma ),\mu _\Sigma )$ and $(M,{\mathcal B}%
(M),\mu _M)$. We have to derive $(\Sigma ,{\mathcal B}(\Sigma ),\mu _\Sigma )$
and $(M,{\mathcal B}(M),\mu _M)$ indirectly from the interpretation of $p(x\
|\pi \ )$ as comming from an interactive probability model 
$({\mathcal P}_\Sigma , {\mathcal P}_M,X,i)$. The absence of an operational definition can
then be justified on principle grounds, but only if the interactive
probabilistic scheme we propose is considered plausible.

\section{Hidden measurements in Hilbert space}

We now turn our attention to the measurement of observables with $n$ distinct outcomes, 
such as the observables related to a spin-n model, 
or to an array of $n$ distinct detectors in a position measurement scheme. 
We will assume that the state space of both the entity and the apparatus is complex Hilbert space.
We start by ascribing a state vector $a \in {\mathcal H}_A$ to the measurement apparatus $A$ and a
state vector $q \in {\mathcal H}_S$ to the system $S$. Next assume there
exists a deterministic interaction $i$ that decides which outcome $x_k$ from
an outcomeset $X=\{x_1,x_2,\ldots,x_n\}$ occurs as a result of an interaction
between the states of the system and the apparatus: 

$$i:{\mathcal H}_S\times {\mathcal H}_A\rightarrow X,\ \ i(q,a)=x $$

\noindent The state of the apparatus, having much more degrees of freedom than the
system it is made to measure, lives in a much bigger Hilbert space, so it is
natural to assume $\dim ({\mathcal H}_A)>>\dim ({\mathcal H}_S)$.
However, all results presented in this article follow if the density of the apparatus 
states is proportional to the area of the subset an n-dimensional subspace where we 
impose the principle of consistent interaction. In the conclusion we briefly touch upon the fact that 
this assumption is a necessity following from an unbiasedness of the apparatus.
Hence for the purpose of the present derivation we need only assume  $\dim ({\mathcal H}_A)$ equals $\dim ({\mathcal H}_S)$.
Having said this, let ${\mathcal H}_n$ denote the set of unit-norm members of an $n$-dimensional Hilbertspace 
over the field of complex numbers, and let $q$ and $a$ belong to this space. 
Hence $i$ is a function: $i:$ ${\mathcal H}_n\times {\mathcal H}_n\rightarrow X$.
Next we connect states to outcomes by means of the concept of an eigenvector.

\begin{definition} 
A set $E=\{e_1,\ldots,e_n\}$ $\subset {\mathcal H}_n$ of $n$
orthogonal vectors is called a set of \emph{eigenvectors} iff  $\forall e_k, e_l \in E$:
\begin{eqnarray*}
\langle e_k,e_l\rangle &=&\delta _{k,l} \\
i(e_k,a) &=&x_k,\forall a\in {\mathcal H}_n
\end{eqnarray*}
\end{definition}

\noindent The vectors $e_k$ play the role of eigenstates for the observable that
corresponds to the measurement being made in the sense that, if the entity
happens to be in one of the states $e_k$, it does not matter with which
apparatus state it interacts; it will always yield the same result and this
result depends only on the eigenstate of the entity. We know from quantum
mechanics the vectors with the desired property ($i(e_k,a)=x_k,\forall a$ ) 
\emph{are} indeed simply the eigenvectors of the self-adjoint operator
corresponding to the relevant observable, but as we did not assume that a
self-adjoint operator represents the measurement of an observable, we have imposed
 this separately. Troughout the rest of the article, indices can take natural values up to $n$ only.

\subsection{The Principle of Consistent Interaction}

To determine the action of $i$, we first define an important set of vectors that we  (in absence of a better name), call
a ``modulus great circle segment'': 
\begin{eqnarray*}
\langle e_k &\circlearrowright &a\rangle \equiv \{c\in {\mathcal H}_n:|c_j|=%
\sqrt{s}|a_j|,\ j\neq k, \\
|c_k| &=&\sqrt{(1-s)+s|a_k|^2},\ 0<s<1\}
\end{eqnarray*}
\noindent So the set of vectors $\langle e_k\circlearrowright a\rangle $ are those
unit vectors in ${\mathcal H}_n$ for which the modulus of each component
equals the downscaled modulus of each component of $c$ (except for $c_j$ )
by a factor $\sqrt{s}$. The last remaining component $c_j$ simply follows
from the normalization requirement on $c$. It is easy to see this is indeed
a segment of a great circle  in the positive $2^n$-tant of $\mathbb{R}^n$, obtained by taking the
modulus of each component of a vector on the unit sphere in $\mathbb{C}^n,$
hence the name ``modulus great circle segment''.

\begin{definition} We say the interaction 
$i:{\mathcal H}_n\times {\mathcal H}_n\rightarrow X$ 
obeys the \emph{\ principle of consistent interaction} (PCI) iff 
$\forall x_k\in X;\ q,a,a^{\prime }\in {\mathcal H}_n, e_r, e_k\in E$:

$$ i(q,a)=x_k\Rightarrow i(q,a^{\prime })=x_k,\forall a^{\prime }\in \langle
e_r\circlearrowright a\rangle ,e_k\neq e_r $$
\end{definition}

\noindent In words, the principle of consistent interaction says that, for a
fixed state $q$ of the entity, if the interaction with an apparatus state $a$
gives rise to an outcome $x_k$, then so does the interaction with any other
apparatus state $a^{\prime }$ that belongs to the modulus great circle
segments between the apparatus state $a$ and any eigenvector belonging to
another outcome than $x_k$. We will first discuss some mathematical
consequences of this principle and postpone a possible interpretation to 
the concluding section of this paper. 
The sets $\langle e_r\circlearrowright a\rangle $ constitute only a
thin subset of the state space ${\mathcal H}_n$ of the apparatus.
Nevertheless, it is evident that the principle poses a severe constraint on
the set of possible partitions of this space. To see just how constraining
the PCI is, let us investigate it by means of the component-wise product of
a complex vector with its complex conjugate, that sends elements of the
complex unit-sphere $S_n=\{z\in \mathbb{C}^n:\sum_{i=1}^nz_iz_i^{*}=1\}$ onto
the $(n-1)$ -simplex 
$\Delta _{n-1}=\{x\in \mathbb{R}_{+}^n:\sum_{i=1}^nx_i=1\}:$
\begin{eqnarray*}
\tau &:&S_n\rightarrow \Delta _{n-1} \\
\tau (z) &=&(z_1z_1^{*},z_2z_2^{*},\ldots,z_nz_n^{*})
\end{eqnarray*}

\noindent Let us translate the PCI to the simplex by means of $\tau$. A simple calculation shows
that $\tau (\langle e_r\circlearrowright a\rangle )=]\tau (e_r),\tau (a)[$,
that is, $\tau $ maps ``modulus great circle segments'' to open
line-segments in $\Delta _{n-1}$. The translation of the PCI to the simplex 
$\Delta _{n-1}$ then reads:
$$i(\tau (q),\tau (a))=x_k\Rightarrow i(\tau (q),\tau (a^{\prime
}))=x_k,\forall \tau (a^{\prime })\in ]\tau (e_r),\tau (a)[, \tau(e_k) \neq \tau(e_r) $$

\noindent It is not difficult to define a partition in the simplex that is consistent
with the PCI. We denote by $]A[$ the relative interior of the convex closure
of $A,$ and define (with slight abuse of the notation $C_k^q$ rather than $C_k^{\tau(q)}$) the sets 

$$ C_k^q=] x_1, \ldots, x_{k-1}, \tau (q), x_{k+1},\ldots, x_n[ $$

\noindent This division of $\Delta _{n-1}$ into separate sets $C_k^q$ takes the form
of a special type of triangulation, which is in a sense a simple
generalization of a barycentric division, and is in fact affinely isomorphic
to it. We have encountered this particular partition in the introduction. 
Just as was the case there, assume now that the interaction $i$ in the simplex is
defined as follows 

$$ \tau(a) \in C_k^q\Rightarrow i(\tau(q),\tau(a))=x_k  $$

\noindent  It is easy to see that for every mapped apparatus state $\tau (a)\in C_k^q$
we indeed have that $]\tau (e_k),\tau (a)[\subset C_k^q$ hence elements of 
$]\tau (e_k),\tau (a)[$ also give rise to an outcome $x_k,$ and as such are
in accordance with the PCI. Likewise, we can see that for every $a\in \tau
^{-1}(C_k^q)$, we have that $\langle e_r\circlearrowright a\rangle \subset
\tau ^{-1}(C_k^q)$ leading to the same conclusion. One can easily convince
oneself intuitively that no other partition of the set of apparatus states
can satisfy the PCI, and refer the interested reader to \cite{AertsS2002}, where a full
proof, utilising mainly elementary convex geometry, can be found. Note that
 $\cup _k\tau ^{-1}(C_k^q)={\mathcal H}_n\backslash M_0,$ where $M_0=\cup
_k\partial [\tau ^{-1}(C_k^q)]$ 
\footnote {It was pointed out to me by T. Durt that the set $M_0$ and the set of points of 
unstable equilibrium in the Bohm-Bub hidden variable model \cite{bohmbub} coincide, showing there is a definite and close relation
between the two approaches}
is the set of boundaries of the closure of
the sets $\tau ^{-1}(C_k^q).$ Clearly $M_0$ is a null set with respect to an 
$n$-measure. Hence, for probabilistic purposes, the 
$\tau^{-1}(C_k^q),k=1,\ldots,n$ constitute what one might call an ``effective
partition'' or a ``partition modulo null-sets'' of the complex unit sphere.

\section{The Born rule}

What constitutes a good measurement? Well, to be sure, a measurement setup
is supposed to give maximal information about the state of the entity it is 
observing and minimal information about the state of the apparatus. For the
probability space ${\mathcal P}_\Sigma $ related to the entity, this means
that $\mu _\Sigma $ becomes a point measure and hence the ensemble $\psi _q$
reduces to a singleton. By a well-known theorem in information theory we
have that, minimization of the information in the probability space $%
{\mathcal P}_M$ related to the apparatus, $\mu _M$ becomes a uniform measure
and the ensemble $\psi _a$ the whole Hilbert space ${\mathcal H}_n$. It turns
out that under these two assumptions, together with the PCI, we recover the
Born rule.

\begin{theorem}: Given an Interactive Probability Model in complex Hilbert
space $({\mathcal P}_{{\mathcal H}_n},{\mathcal P}_{{\mathcal H}_n},X,i)$ with $%
i $ satisfying the PCI. Assume the preparation ${\mathcal B}(\Sigma )\times 
{\mathcal B}(M)\ni \pi =(q,{\mathcal H}_n)$ where $q$ is a singleton. With $%
\{e_1,\ldots,e_n\}$ a set of eigenvectors and $e_k$ the eigenvector
corresponding to the outcome $x_k\in X,$ we have: 
$$p(x_k\ |\pi \ )=|\langle q,e_k\rangle |^2 $$
\end{theorem}

\medskip \textbf{Proof:} We start with the definition of the interactive
probability under the assumptions of the theorem: 
\begin{eqnarray*}
p(x_k\ |\pi \ ) &=&\frac 1{\rho (q,{\mathcal H}_n)}\int_{i^{-1}(x_k)}d\rho \\
&=&\frac{\mu _\Sigma (q)}{\mu _\Sigma (q)\mu _M({\mathcal H}_n)}\int_{\tau
^{-1}(C_k^q)}d\mu _M \\
&=&\frac{\mu _M(\tau ^{-1}(C_k^q))}{\mu _M({\mathcal H}_n)}
\end{eqnarray*}

\noindent This last equation simply tells us that the probability of getting the
outcome $x_k$ equals the ratio of the apparatus states that result in that
outcome to the total of all possible apparatus states. The calculation of
the quantity $\mu _M(\tau ^{-1}(C_k^q))$ is greatly facilitated by realizing 
$\tau$ preserves probability measures. In virtue of a lemma
presented after this argument, the last expression becomes: 
$$
\frac{\nu (C_k^q)}{\nu (\Delta _{n-1})} 
$$
The calculation of this last quantity was outlined in the
introduction of this article and demonstrated explicitely in \cite{Aerts1986}. 
\begin{eqnarray*}
&=&\tau (q)\cdot \tau (e_k)=q_kq_k^{*} \\
&=&|\langle q,e_k\rangle |^2
\end{eqnarray*}

\begin{lemma}Let $(\Delta _{n-1},{\mathcal B}(\Delta _{n-1}),\mu )$ and $%
(S_n,{\mathcal B}(S_n),\nu )$ be two measure spaces. Then for $A\in $ $%
{\mathcal B}(\Delta _{n-1})$ and $\tau ^{-1}(A)\in {\mathcal B}(S_n)$, we have:

$$\nu (\tau ^{-1}(A))=\frac{2\pi ^n}{\sqrt{n}}\mu (A) $$
\end{lemma}

\textbf{Proof:} Let $A$ be an arbitrary open convex set in $\Delta _1$: $%
A=\{(x_1,x_2):a<x_1<b,\ x_2=1-x_1\}$. Evidently, $\mu (A)=\sqrt{2}(b-a)$.
Let $B$ be the pull-back of $A$ under $\tau :$ 
\begin{eqnarray*}
B &=&\{(z_1,z_2)\in Z_1\times Z_2\subset \mathbb{C}^2:Z_1=\{z_1:a<|z_1|^2<b\},
\\
Z_2 &=&\{z_2:\ z_2=\sqrt{1-|z_1|^2}e^{i\theta },\theta \in [0,2\pi [\}\}
\end{eqnarray*}

Clearly,  $$ \nu (B)=\nu (Z_1)\nu (Z_2)=\pi (b-a).2\pi =\frac{2\pi ^2}{\sqrt{2}}\mu (A) $$
\noindent Hence the theorem holds for convex sets if $n=2$. This conclusion can
readily be extended to an arbitrary $(n-1)$-dimensional rectangleset $A$ in $%
\Delta _{n-1}:$

$$ A=\{(x_1,\ldots,x_{n-1},1- \sum_{i=1}^{n-1} x_i): \forall i=1,\ldots, n-1:
a_i<x_i<b_i; \ a_i,b_i \in [0,1]\}  $$

Its measure factorizes into: 

$$ \mu (A)=\sqrt{n}\prod_{i=1}^{n-1}(b_i-a_i)  $$
Next consider n-tuples of complex numbers: 
\begin{eqnarray*}
B &=&\{(z_1,z_2,\ldots,z_n)\in Z_1\times \ldots\times Z_n\} \\
Z_i &=&\{z_i \in \mathbb{C}: a_i<|z_i|^2<b_i , i \neq n \},\  \\
z_n &=&\sqrt{1-|z_1|^2-\ldots-|z_{n-1}|^2}e^{i\theta _n},\theta _n\in [0,2\pi
[\}\}
\end{eqnarray*}
Clearly $\tau (B)=A$. The measure of $B$ can be factorized as:

\begin{eqnarray*}
\nu (B) &=&\nu (Z_1)\nu (Z_2)\ldots\nu (Z_n) \\
&=&2\pi \prod_{i=1}^{n-1}\pi (b_i-a_i)=\frac{2\pi ^n}{\sqrt{n}}\mu (A)
\end{eqnarray*}
\noindent Hence the theorem holds for an arbitrary rectangleset $A$. But every open
set in $\Delta _{n-1}$ can be written as a pairwise disjoint countable union
of rectangular sets. It follows that $\nu (\tau ^{-1}(\cdot ))=\frac{2\pi ^n%
}{\sqrt{n}}\mu (\cdot )$ for all open sets in $\Delta _{n-1}$. Both $\nu $
and $\mu $ are finite Borel measures because $\Delta _{n-1}$ and $S_n$ are
both compact subsets of a vectorspace of countable dimension. Therefore they
must be regular measures. But a regular measure is completely defined by its
behavior on open sets. Hence the theorem holds for Borel sets.

\section{Concluding Remarks}

\noindent Besides the fact that the PCI defines the partition of the set of apparatus states,
it is also interpretable as some sort of ``proposal-consistent-answer-game''.
To see this, make the comparison  with the well-known game of ``warm'' and ``cold''.
The object of the game is to guess the location of an unknown object in a room, using the cluess ``warmer'' and
``colder'' given by someone who knows the location of the object. 
The equivalent of the PCI for this game would be that if the guesser his next guess 
is  \emph{further} from the object than a former guess, his reply has to be ``colder''.
Imagine now a straightforward multi-dimensional generalization of the game played in the 
$(n-1)$-simplex, and with as possible answers the $n$ vertices of the simplex. 
The state vector $a$ then, represents the measurement apparatus and is a 
``proposal'' both \emph{to} the state and \emph{for} the state, as if the 
measurement asks the question: can you give me a clue about your true location 
if my guess would be that it is somewhere here you are residing? 
Now the entity, in response to that proposal has to give a hint about its true location by giving the unique outcome that is in accordance with the PCI.
This answer can only be one of the $n$ outcomes corresponding to the eigenvectors and, seen from the point of view of the guesser
(the apparatus), it gives $n$ alternative directions to choose from. The PCI does not tell what happens when the next guess is closer to 
the eigenvector corresponding to the outcome given to the former guess. It doesn't need to.
What the PCI requires, is that the response of the entity is such that if the
answer was ``vertex $x_i$'' and the guesser chooses to ignore that directional hint and places his guess
closer in the direction of another vertex (rather than closer to vertex $x_i$) the answer will still have to be $x_i$.
Once you think of it this way, the PCI indeed expresses a very basic form of consistency, and it is nice to see
 that this condition alone partitions the set of apparatus states. As such the PCI seems to show a relationship between the geometry of Hilbert space
and the probabilistic inferences made therein. The model we propose in the case of quantum
mechanics differs from the game of warm and cold in that each new interaction forgets
the outcome that was given in response to interaction with a former apparatus
state. This is connected to the minimization of the information regarding the interactional part of the apparatus state and the corresponding uniform density of states. 
It is also essentially a matter of the apparatus being unbiased: the apparatus
should not be more sensitive to some states than to others, nor should it 
know in advance which entity state is going to be presented to it. Together, these assumptions led us to the Born rule.
Mathematically speaking, Gleason's theorem gives you more for less, apparently rendering the result redundant. However, we have gained an interpretation. 
If there is in nature something like "the observer and the observed together producing the phenomenon", then we believe the scheme outlined here is sensible and, 
as we hope to have shown, not too dificult to translate to complex Hilbert space to recover the Born rule by an integration over unknown observer states, 
in accordance with the original hidden measurement proposal.

\end{document}